\documentclass{article}
\usepackage{epsfig}
\begin{document}


\def\DAF{DA$\Phi$NE }
\def\Repsp{Re($\epsilon'/\epsilon $)}
\def\epsp{$\epsilon'/\epsilon$}
\def\ppm{$\pi^{+}\pi^{-}$}
\def\p00{$\pi^{0}\pi^{0}$}
\def\KL{K$^{0}_{L}$}
\def\KS{K$^{0}_{S}$}
\def\KKbar{$K^{0}\overline{K^{0}}$}
\def\K3{$K_{l3}$}
\def\K4{$K_{l4}$}
\def\lum{cm$^{-2}$s$^{-1}$}
\def\K00{K$^{0}_{S} \rightarrow \pi^{0}\pi^{0}$}
\def\Kpm{K$^{0}_{S} \rightarrow \pi^{+}\pi^{-}$}
\def\Lzz{K$^{0}_{L} \rightarrow \pi^{0}\pi^{0}$}
\def\Lpm{K$^{0}_{L} \rightarrow \pi^{+}\pi^{-}$}
\def\L3p{K$^{0}_{L} \rightarrow \pi^{+}\pi^{-}\pi^{0}$}
\def\Len{K$^{0}_{L} \rightarrow \pi^{\pm}e^{\mp}\nu$}
\def\Sep{K$^{0}_{S} \rightarrow \pi^{\pm}e^{\mp}\nu$}
\def\nnn{K$^{0}_{L} \rightarrow \pi^{0}\pi^{0}\pi^{0}$}
\def\Ps0{P$_{S0}$}
\def\Pl1{P$_{L1}$}
\def\epm{e$^{+}$e$^{-}$}

\begin{flushright}
 \small
 Contributed paper to {\it Lepton Photon 2001}\\
 \small
 Rome, July 23-28 2001 
\end{flushright}
\vspace*{0.5truecm}
\begin{center}
{\LARGE Studies of \KS\ decays with the KLOE \\ 
           detector at \DAF \\}
\end{center}
\vspace{0.5cm}
\begin{center}
{\large \bf The KLOE Collaboration:} 
\end{center}
\vspace{0.3cm}
\begin{center}
A.~Aloisio$^g$, F.~Ambrosino$^g$, 
A.~Antonelli$^c$, M.~Antonelli$^c$, 
C.~Bacci$^l$, G.~Barbiellini$^n$, 
F.~Bellini$^l$
G.~Bencivenni$^c$, S.~Bertolucci$^c$, C.~Bini$^j$
% ...

\vspace{0.5cm}
{\small $^a$Dipartimento di Fisica dell'Universit\`a e Sezione INFN, Bari, Italy}\\
{\small $^b$Institute of High Energy Physics of Academica Sinica, Beijing, China}\\
{\small $^c$Laboratori Nazionali di Frascati dell'INFN, Frascati, Italy}\\
{\small $^d$Institut f\"ur Experimentelle Kernphysik, Universit\"at Karlsruhe, Germany}\\
{\small $^e$Dipartimento di Fisica dell'Universit\`a e Sezione INFN, Lecce, Italy}\\
{\small $^f$Institute for Theoretical and Experimental Physics, Moscow, Russia}\\
{\small $^g$Dipartimento di Scienze Fisiche dell'Universit\`a e Sezione INFN, Napoli, Italy}\\
{\small $^h$Dipartimento di Fisica dell'Universit\`a e Sezione INFN, Pisa, Italy}\\
{\small $^j$Dipartimento di Fisica dell'Universit\`a ``La Sapienza'' e Sezione INFN, Roma, Italy}\\
{\small $^k$Dipartimento di Fisica dell'Universit\`a ``Tor Vergata'' e Sezione INFN, Roma, Italy}\\
{\small $^l$Dipartimento di Fisica dell'Universit\`a ``Roma Tre'' e Sezione INFN, Roma, Italy}\\
{\small $^m$Physics Department, State University of New York at Stony Brook, USA}\\
{\small $^n$Dipartimento di Fisica dell'Universit\`a e Sezione INFN, Trieste, Italy}\\
{\small $^o$Physics Department, University of Virginia, USA}
\end{center}


\abstract{
The KLOE detector at \DAF, the Frascati $\phi$-factory, has collected
about 30 pb$^{-1}$ by the end of year 2000. The $\phi$(1020) meson decays
about 34$\%$ of the times into a \KL-\KS\ pair; \DAF is therefore an
exceptional source of almost monochromatic, tagged \KS\ particles,
allowing for detailed studies of their more rare decays. 
The above mentioned integrated luminosity corresponds to about 30 millions
produced \KS. 

In KLOE the presence of a \KS\ is easily tagged by the observation of the
corresponding \KL\ impinging onto the electromagnetic calorimeter before decay.
Thanks to this technique we have performed a preliminary measurement of the ratio 
of the partial decay 
widths of the \KS\ into two charged and two neutral pions:
{\bf B(\Kpm)/B(\K00)} = 
  2.23 $\times$ (1 $\pm$ 0.35$\times$ 10$^{-2}$ (stat) $\pm$ 1.5$\times$ 10$^{-2}$ (syst) ). 

We have also observed several hundreths semileptonic decays of the \KS\, 
by far the largest sample of this decay mode ever observed by any experiment
so far. The preliminary estimate of the corresponding branching ratio is 
{\bf B(\Sep)} = (6.8 $\pm$ 0.3 (stat))$\times$10$^{-4}$; systematic effects are 
still under study and are presently estimated to be at the few percent level. }

\vskip 1.cm

\section{INTRODUCTION}
The KLOE detector at \DAF, the Frascati $\phi$-factory, has started physics
data taking in April 1999. It has collected about 30 pb$^{-1}$ by the end of 
 year 2000. 

The $\phi$(1020) meson decays
$\sim$ 34$\%$ of the times into a \KL-\KS\ pair; at peak energy, 
about 1 million of such decays occur every delivered pb$^{-1}$. 
\DAF is therefore an
exceptional source of almost monochromatic, tagged \KS\ particles,
allowing for detailed studies of their more rare decays.

In the present paper, the status of the analysis about two different
\KS\ decay channels is presented.

 Firstly, a measurement of the ratio
among the branching ratios into two charged and neutral pions is presented. This 
is relevant for CP violation studies, since it enters the double ratio
from which \Repsp\ is derived. Moreover it is of interest for chiral 
perturbation theory studies, especially
if the radiation of soft photons in the charged decay is properly taken
into account. 

Secondly, a measurement of the branching ratio of the decay \Sep\ is presented. 
Up to now, only one measurement of this branching ratio exists, based 
on a data sample of 75 events \cite{semnov}. In the present analysis 
the measurement is performed using a sample of about 600 event candidates 
with a background contamination of less than 10$\%$.
   
\section{THE KLOE DETECTOR}
The KLOE (KLOngExperiment) detector~\cite{kp}, designed  with the 
primary goal of measuring \epsp\ with a sensitivity of the order of one 
part in ten thousand, is also particularly well suited to perform studies on all
charged and neutral decays of the \KS\ meson.

\begin{center}
\begin{figure}
\vspace{9pt}
\hspace{50pt}
\epsfig{figure=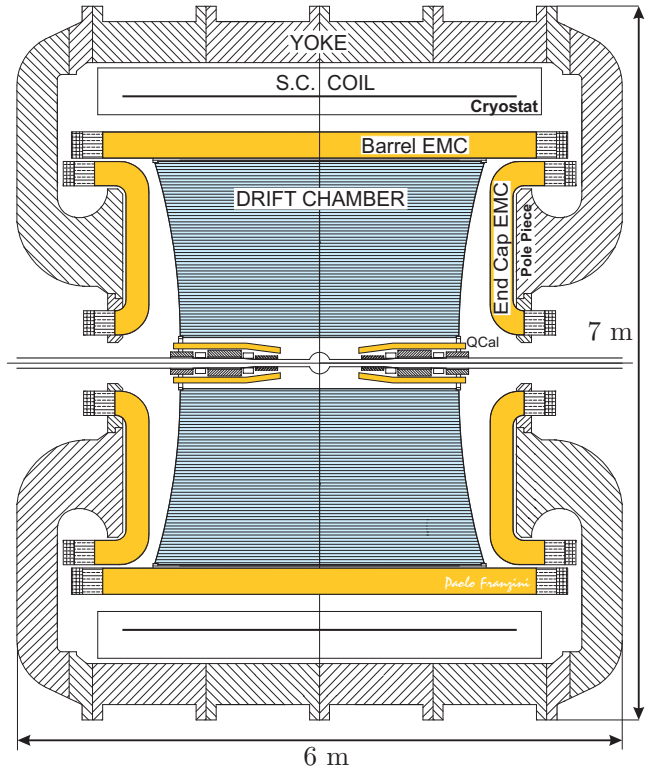,width=8.5cm}
\caption{Side view of the KLOE detector.}
\label{k1}
\end{figure}
\end{center}

It consists of a large tracking chamber, a hermetic electromagnetic 
calorimeter and a large magnet surrounding the whole detector, consisting
of a superconducting coil and an iron yoke (see figure \ref{k1}). 

The tracking chamber \cite{dcp,dpap} (DC) is a cylindrical, 2 m radius,
 3.3 m long drift chamber. 
The total number of wires is 52140, out of which 12582 are the sense ones.
It operates with a low-Z, He gas mixture, to minimize multiple scattering of
charged particles and
regeneration of \KL 's. The 58 concentric layers of wires are strung in 
an all-stereo geometry, with constant inward radial displacement at the 
chamber center. A spatial resolution better than 200 $\mu$m is obtained. 
The momentum resolution  for 510 MeV/c electrons and positrons is 1.3 MeV/c, 
in the angular range 130$^{\circ} > \theta >$ 50$^{\circ}$.
 
The electromagnetic calorimeter\cite{tp,epap} (EmC) is a lead scintillating 
fibers sampling 
calorimeter, divided into a barrel section and two endcaps. The modules of
 both sections
are read out at the two ends by a total of 4880 photomultipliers. 
In order to minimize
dead zones in the overlap region between barrel and endcaps, the modules 
of the latter are bent outwards with respect to the decay region. 

The calorimeter was designed to detect with very high efficiency  photons 
with energy as low as 20 MeV, and to accurately measure their energy and time of flight.
Absolute calibrations of energy and time scales are performed using collision 
data. 
An energy resolution of 5.7$\%$/$\sqrt{E(GeV)}$ is achieved throughout the 
whole calorimeter together with a linearity in energy response better than
1$\%$ above 80 MeV and 4$\%$ between 20 to 80 MeV. 

Moreover, $\gamma$
samples from different processes are selected to measure the time resolution
at various energies; it scales according to the law 
$\sigma_{t}$=(54/$\sqrt(E(GeV) \oplus$ 147) ps, where the first term is in 
agreement with test beam data, while the second, to be added in quadrature,
is dominated by the intrinsic time spread due to the bunch length. 

\section{STUDIES ON PHYSICS CHANNELS}

\subsection{Tagging of \KS\ decays}
When a $\phi$ meson decays into two neutral kaons $C$-parity 
invariance forces the two kaons to be in a
\KS-\KL\ state. The observation of a \KS, therefore, {\it tags} the presence of
the \KL\ in the opposite hemisphere. 
Similarly, \KS\ decays can be selected by observing the \KL\ on the other
side.
 
As \KS\ tagging strategy, one can either look for a
charged vertex well inside the DC volume, or identify a EmC cluster compatible 
with being due to a slowly moving ($\beta \approx$ 0.22) neutral particle (so called 
'KCRASH' event). Actually, more than one half of the \KL 's reach 
the calorimeter before they decay. For the above reason the 'KCRASH' 
tag  provides a particularly clean, high statistics \KS\ sample. 
 
More specifically, events are selected on the basis of the two following requests: 
\begin{enumerate}
\item{The presence of a EmC cluster with energy larger than 50 MeV, and 
transverse radius larger than 60 cm, due to the \KS\ decay; it is needed to 
determine the 
t$_{0}$ of the event, i.e. the time at which the $\phi$ production and 
decay occurred.}
\item{The presence of a EmC cluster in the barrel region with energy larger 
than 100 MeV and time compatible with being due to a particle moving at 
a velocity in the $\phi$ rest frame  0.195 $< \beta^{\ast} <$ 
 0.2475 (the KCRASH).}
\end{enumerate}

\begin{figure}
\hspace{20pt}
\epsfig{figure=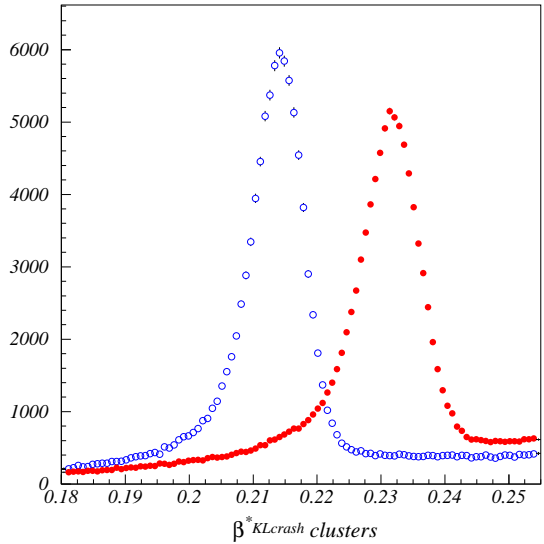,width=8.5cm}
\caption{$\beta^{\ast}$ distributions for KCRASH clusters when the accompanying 
\KS\ decays into two charged (open circles) or neutral (black circles) pions.}
\label{fbeta}
\end{figure}

The tag efficiency is slightly dependent on the \KS\ decay type, since the time zero
estimate (first point above) is determined by particles with different 
velocities (prompt photons in the case of \K00\ events, pions in \Kpm\ ones, pions
or electrons for semileptonic decays). 
For instance, the distributions for the reconstructed
$\beta^{\ast}$ for charged and neutral two pions decays are shown in figure \ref{fbeta};
it turns out that  
the ratio of the efficiencies for having a KCRASH in the above mentioned velocity 
interval is $\epsilon^{+-}$/$\epsilon^{00}$ = (95.030 $\pm$ 0.005)$\%$, where the error
is statistical only. 

In the following, all events are tagged making use of the KCRASH prescription. 

\subsection{\Kpm\ and \K00\ decays}
\begin{figure}
\vspace{9pt}
\hspace{20pt}
\epsfig{figure=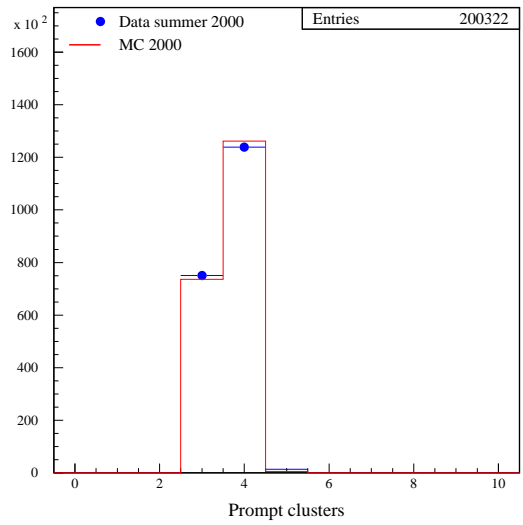,width=9.5cm}
\caption{Distribution of the number of prompt clusters, N$_{\gamma}$
 found in KCRASH events for data (black points) and Monte Carlo (dashed line).
The request N$_{\gamma} >$ 2 is applied, to reject machine background events.}
\label{cprom}
\end{figure}

\begin{figure}
\hspace{20pt}
\epsfig{figure=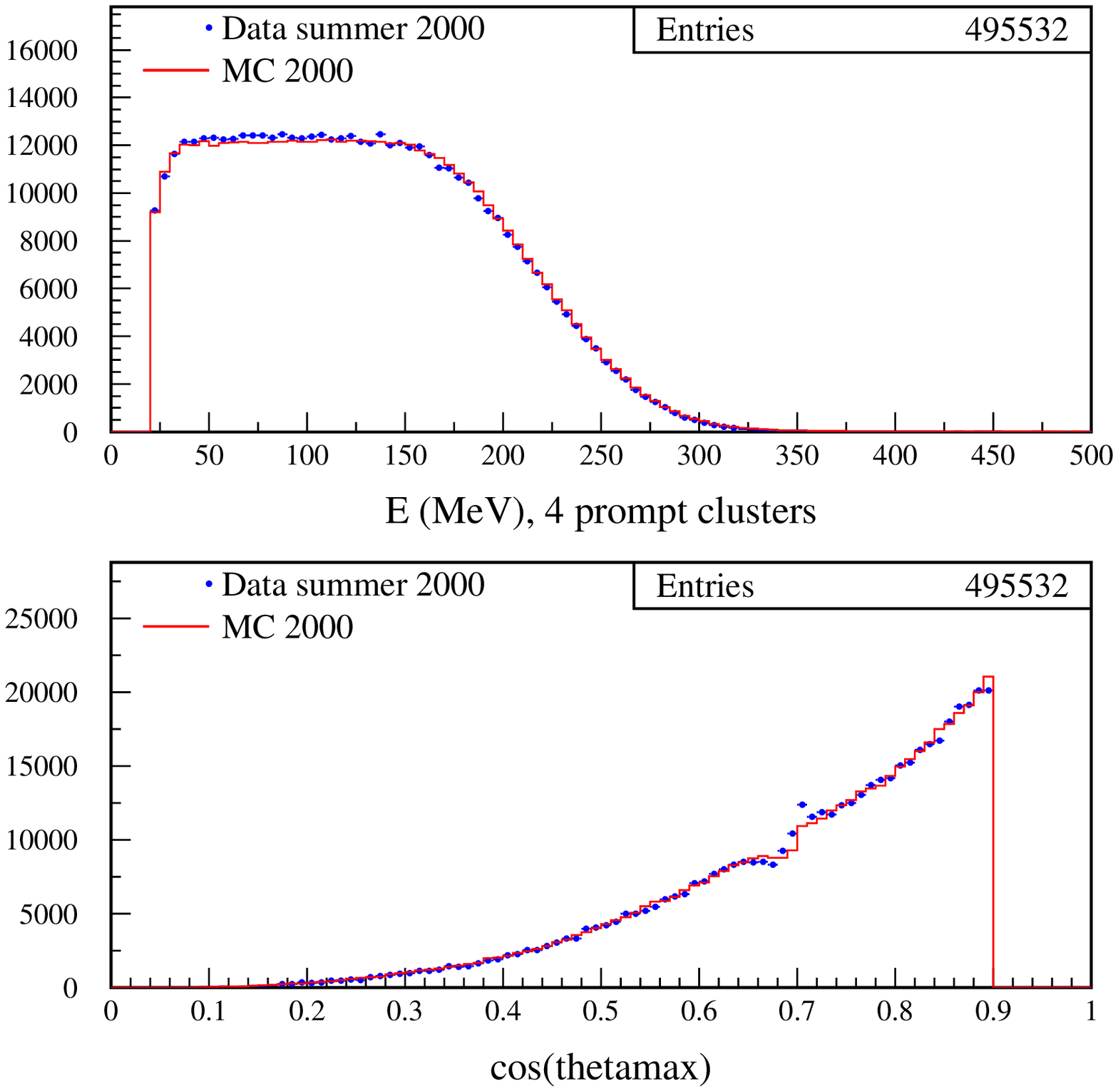,width=9.5 cm}
\caption{Energy (upper plot) and angular (lower plot) distributions of the photons
 in four $\gamma$ events. Points are data, solid line is the Monte Carlo prediction
 for \K00\ events.}
\label{pdist}
\end{figure}

The \KS\ decays into two  neutral pions are selected requiring  
the presence of four EmC clusters with a timing compatible with the hypothesis 
of being due to prompt photons (within 5 $\sigma$'s), and energy larger than 
20 MeV. The prompt clusters distribution for the data taken during summer 2000
is shown in figure \ref{cprom} together with the Monte Carlo expectation. The distribution
agree well between each other. The energy spectrum and the angular distribution 
for the photons of the events with four prompt clusters are shown in figure \ref{pdist}.
Again, good agreement between data and Monte Carlo is observed.  

Photon detection efficiency is estimated by real data using $\gamma$'s 
in the decays $\phi \rightarrow \pi^{+}\pi^{-}\pi^{0}$ as a control sample. 
The final selection efficiency for the \K00\ decay channel is 
$\epsilon_{00}$=(56.7$\pm$0.1)$\%$, dominated by acceptance. 

\begin{figure}
\vspace{9pt}
\hspace{20pt}
\epsfig{figure=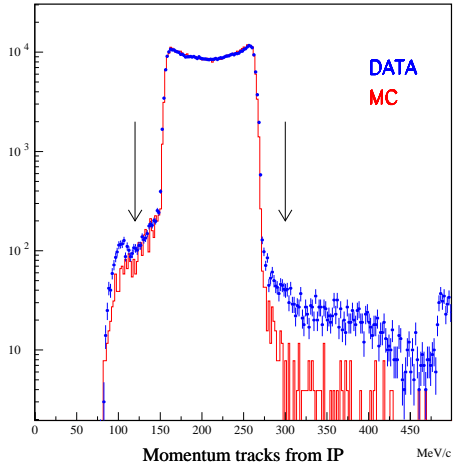,width=8.5cm}
\caption{Momentum distribution for the tracks originating from the I.P. in KCRASH
selected events. Black points are data, while the solid histogram is the Monte Carlo 
expectation for \Kpm\ events. Tracks with P$>$ 300 MeV/c are mostly due to machine 
background, while the peak at P$\sim$ 100 MeV/c is due to the residual contamination
from $\phi \rightarrow K^{+}K^{-}$ events.}
\label{trch}
\end{figure}

The selection of \Kpm\ events proceeds through the request of two oppositely
charged tracks with polar angle  in the interval 
30$^{\circ} < \theta <$ 150$^{\circ}$, originating in a cylinder of 4 cm radius and 
10 cm length around the interaction point. 
A further request is applied on the measured momenta to remove the
residual background due to charged kaon decays: 120 $<$ p(MeV/c) $<$ 300 (see
figure \ref{trch}). Both tracks are also required to impinge to the calorimeter, 
in order to enhance the probability for having a good t$_{0}$ determination. 

The track reconstruction efficiency is measured in momentum and polar angle 
bins from data subsamples. The final selection efficiency is 
$\epsilon_{+-}$=(58.5$\pm$0.1)$\%$, again dominated by acceptance.

The trigger efficiency is determined with real data for both decay 
types. It is (99.69 $\pm$ 0.03)$\%$ for the neutral decay and 
(96.5 $\pm$ 0.1)$\%$ for the charged one. The above figure includes also 
the probability for having at least one good cluster to determine the 
t$_{0}$ of the event, as explained in the previous paragraph. 

Background levels are kept well below 1$\%$ for both decay types. 

Using part of the data acquired in year 2000, corresponding to $\sim$10 pb$^{-1}$, 
872748 \Kpm\ and 414118 \K00\ decays have been selected, providing:
{\bf B(\Kpm)/B(\K00)} = 
  2.23 $\times$ (1 $\pm$ 0.35$\times$ 10$^{-2}$ (stat) $\pm$ 1.5$\times$ 10$^{-2}$ (syst) )

to be compared with the present PDG value \cite{pdg} 
2.197 $\times$ (1 $\pm$ 1.2$\times$ 10$^{-2}$ (stat) 
$\pm$ 1.5$\times$ 10$^{-2}$ (syst) ). 

Systematics are dominated by 
residual uncertainties in photon counting and in the understanding of the 
difference between the tagging efficiencies for the two channels. 
More precise studies are presently under way.

\subsection{\Sep\ decays}
In order to search for \Sep\ decay candidates, events with a KCRASH and two  
oppositely charged tracks from the interaction region are initially selected.
Events are then rejected if the two tracks 
invariant mass (in the pion hypothesis) and the resulting \KS\ momentum 
in the $\phi$ rest frame  are compatible with those expected for a \Kpm\ decay.   
According to Monte Carlo, this preselection has an efficiency, after the tag, of
 $\sim$ 62.4$\%$ on the signal.

In order to perform a time of flight identification of the charged particles, 
both tracks are required to be associated with a EmC cluster. The acceptance for
such request, estimated by Monte Carlo, is (51.1 $\pm$ 0.2) $\%$. The time of flight
difference $\Delta\delta$t for the two charged particles in both e-$\pi$ 
and $\pi$-$\pi$ hypotheses is then computed; events are accepted if 
$| \Delta\delta$t($\pi$-$\pi $)$| >$ 1.5 ns  and $ | \Delta\delta$t($\pi$-e)$| <$ 1 ns
and $ | \Delta\delta$t(e-$\pi$)$| >$ 3 ns. 
The efficiency on the signal, estimated by means of \KL 's decaying into 
$\pi$e$\nu$ before the DC internal wall, is (82.0$\pm$0.7)$\%$ .  

Also the trigger efficiency as well as the one for correctly associating a track 
to a cluster and 
for having a good t$_{0}$ determination is measured directly on data, 
making use of \Len, $\phi \rightarrow \pi^{+}\pi^{-}\pi^{0}$ and \Kpm subsamples. 
The product of these efficiencies turns out to be (81.7 $\pm$ 0.5) $\%$. 

\begin{figure}
\vspace{9pt}
\hspace{20pt}
\epsfig{figure=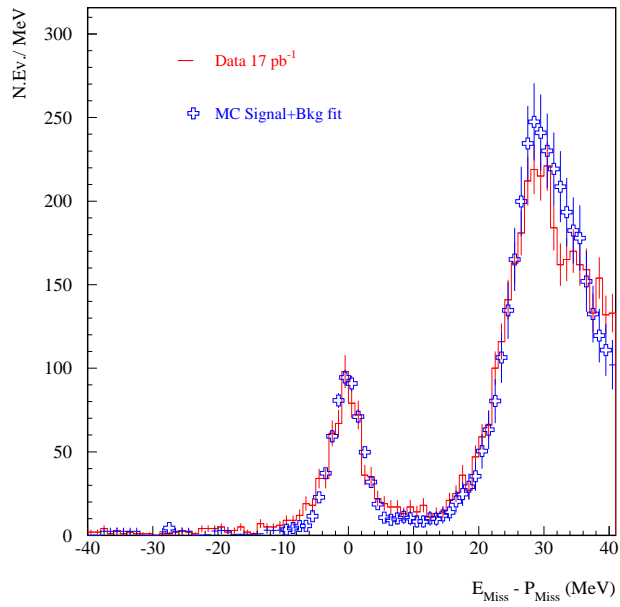,width=9.5cm}
\caption{Distribution of the difference between missing energy and missing momentum
for \Sep\ candidates. The peak at zero is the signal. The distribution is fit to 
the Monte Carlo of signal and background in the range -40 MeV + 40 MeV.}
\label{fitse}
\end{figure}

The event is finally kinematically closed.
The \KS\ momentum is estimated making use of the measured direction of the \KL\
and of the $\phi$ 4-momentum. The missing energy and momentum of the 
\KS\ -$\pi$-e system, corresponding to the neutrino's ones, 
are then computed. Their difference is distributed as
 in figure \ref{fitse}; it must be equal zero for the signal. Data are fit
using MC spectra for both signal and the residual background, due mostly to
\Kpm\ events with an early decay of one of the two pions.
    
Using data corresponding to a luminosity of $\sim$ 17 pb$^{-1}$, 
the measured yield is N(\Sep) = 627 $\pm$ 30 events, for a total efficieny of 
(21.8 $\pm$ 0.3) $\%$. 
The total number of events is then normalised to the amount of observed \Kpm\ 
events, to give {\bf B(\Sep)} = (6.8 $\pm$ 0.3 (stat))$\times$10$^{-4}$. 
In the ratio, the tagging efficiency, which is the largest cause of 
systematic uncertainty, cancels out identically.
Other systematic effects, presently under study, are preliminarly estimated to be 
at a few percent level.

This result can be compared with the one obtained by the CMD2 Collaboration\cite{semnov}: 
B(\Sep) = (7.2 $\pm$ 1.2)$\times$10$^{-4}$, and with the prediction obtained assuming
$\Gamma_{S}$=$\Gamma_{L}$ : B(\Sep) = (6.70 $\pm$ 0.07)$\times$10$^{-4}$. 

\end{document}